\documentclass[letterpaper]{article} % DO NOT CHANGE THIS
\usepackage{aaai2026}  % DO NOT CHANGE THIS
\usepackage{times}  % DO NOT CHANGE THIS
\usepackage{helvet}  % DO NOT CHANGE THIS
\usepackage{courier}  % DO NOT CHANGE THIS
\usepackage[hyphens]{url}  % DO NOT CHANGE THIS
\usepackage{graphicx} % DO NOT CHANGE THIS
\usepackage{natbib}  % DO NOT CHANGE THIS AND DO NOT ADD ANY OPTIONS TO IT
\usepackage{caption} % DO NOT CHANGE THIS AND DO NOT ADD ANY OPTIONS TO IT
\usepackage{algorithm}
\usepackage{algorithmic}

\usepackage{amssymb}
\usepackage{booktabs}

\usepackage{multirow}
\usepackage{subcaption} % for subfigures
\usepackage{enumitem}
\usepackage{amsmath} 

\usepackage{newfloat}
\usepackage{listings}
\DeclareCaptionStyle{ruled}{labelfont=normalfont,labelsep=colon,strut=off} % DO NOT CHANGE THIS
\floatstyle{ruled}
\newfloat{listing}{tb}{lst}{}
\floatname{listing}{Listing}
\title{Semantic Item Graph Enhancement for Multimodal Recommendation}
\author{
    Xiaoxiong Zhang, Xin Zhou, Zhiwei Zeng, Dusit Niyato, Zhiqi Shen\\
    College of Computing and Data Science, Nanyang Technological University, Singapore\\
    zhan0552@e.ntu.edu.sg,
    \{xin.zhou, zhiwei.zeng, dniyato, zqshen\}@ntu.edu.sg
}

\usepackage{bibentry}
\begin{document}

\maketitle

\begin{abstract}
Multimodal recommendation systems have attracted increasing attention for their improved performance by leveraging items' multimodal information. Prior methods often build modality-specific item-item semantic graphs from raw modality features and use them as supplementary structures alongside the user-item interaction graph to enhance user preference learning. However, these semantic graphs suffer from semantic deficiencies, including (1) insufficient modeling of collaborative signals among items and (2) structural distortions introduced by noise in raw modality features, ultimately compromising performance. To address these issues, we first extract collaborative signals from the interaction graph and infuse them into each modality-specific item semantic graph to enhance semantic modeling. Then, we design a modulus-based personalized embedding perturbation mechanism that injects perturbations with modulus-guided personalized intensity into embeddings to generate contrastive views. This enables the model to learn noise-robust representations through contrastive learning, thereby reducing the effect of structural noise in semantic graphs. Besides, we propose a dual representation alignment mechanism that first aligns multiple semantic representations via a designed Anchor-based InfoNCE loss using behavior representations as anchors, and then aligns behavior representations with the fused semantics by standard InfoNCE, to ensure representation consistency. Extensive experiments on four benchmark datasets validate the effectiveness of our framework.
\end{abstract}

% Finally, we propose a dual representation alignment mechanism to effectively align and fuse behavior and semantic representations. It first aligns multiple semantic representations using an Anchor-based InfoNCE loss, where behavior representations serve as anchors, and then aligns the behavior representations with the fused semantic representations.


%     \link{Code}{https://anonymous.4open.science/r/SIGER-CB34}

\section{Introduction}

A recommendation system is designed to model user preferences and deliver tailored item suggestions \cite{zhou2024advancing, chen2019collaborative}.
As a specific branch of recommendation systems, \textbf{M}ulti-\textbf{M}odal \textbf{R}ecommendation systems (MMR) incorporate diverse item modalities, such as visual and textual features, to more comprehensively capture user preferences and improve recommendation accuracy \cite{liu2022multimodal, xu2021multi, cm32025arxiv, song2023mm, LIRD2025arxiv, bu2010music}.
This line of research has recently gained growing attention.

A widely adopted research paradigm in MMR involves constructing modality-specific item-item semantic graphs from raw modality features and employing them as auxiliary structures to the user-item interaction graph to improve user preference modeling, as exemplified by LATTICE \cite{lattice}, MICRO \cite{micro}, MGCN \cite{mgcn}, FREEDOM \cite{freedom}, DA-MRS \cite{xv2024improving}, etc. These methods typically construct item-item semantic links by computing the cosine similarity between raw modality features, connecting each item to its Top-K nearest neighbors (commonly, K = 10). Graph convolution operations are applied separately to the semantic and user-item graphs to learn semantic and behavioral representations, which are then fused to form the final representations for recommendation. Despite certain progress, the effectiveness of this approach is still limited by the inherent semantic deficiency of the constructed item-item graphs. 
\vspace{-1em}
\begin{figure}[H]
  \centering
  \begin{minipage}{0.49\linewidth}
      \centering
      \includegraphics[width=\linewidth]{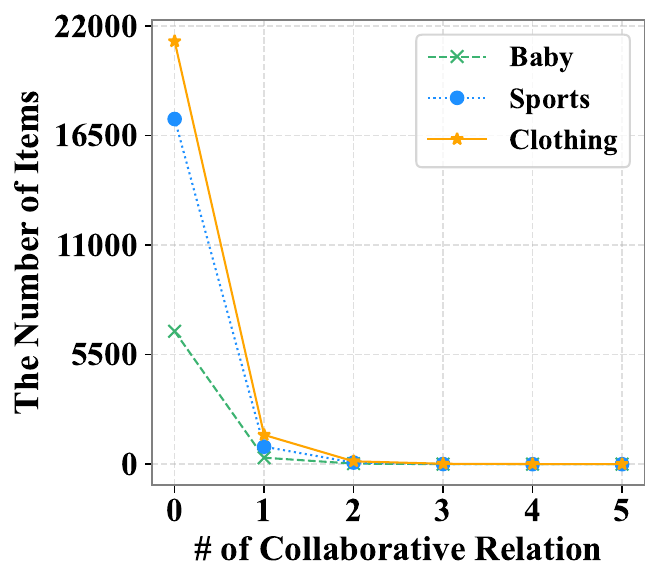}
      \subfloat{(a) Image Modality}
  \end{minipage}
  \hfill
  \begin{minipage}{0.49\linewidth}
    \centering
      \includegraphics[width=\linewidth]{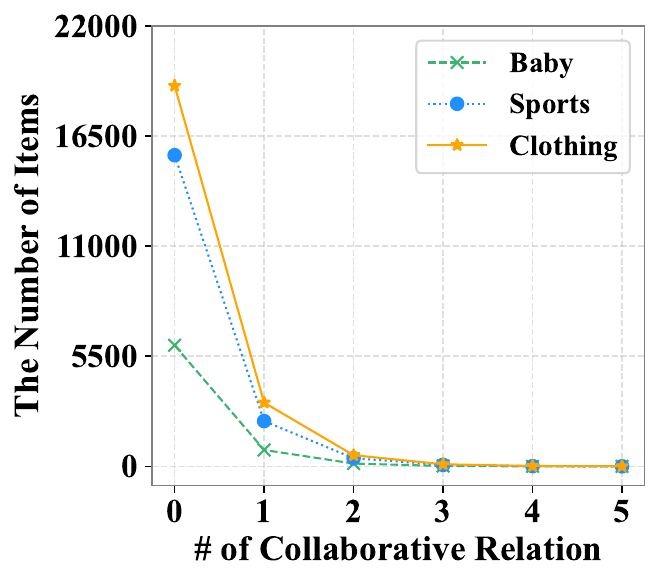}
      \subfloat{(b) Text Modality}
  \end{minipage}
  \caption{The distribution of captured collaborative items by Modality-Specific Item-Item Semantic Graphs. For the three datasets, most items are linked to none or one of collaborative items within the modality-specific item-item graphs.}
  \label{fig:intro}
\end{figure}
\vspace{-1em}
On one hand, the constructed modality-specific item-item semantic graphs typically capture modality-aware similarity relations but fail to effectively capture collaborative (i.e., co-purchasing) relations. As a result, encoding on such graphs primarily strengthens representation proximity among modality-similar items while weakening that among co-purchased ones, limiting the model’s capacity to further improve performance. We quantitatively evaluate the extent to which these semantic graphs capture collaborative relations. Specifically, we retrieve each item’s top-5 co-occurring items from the interaction graph and evaluate how many are present in its modality-specific semantic neighbors. Figure \ref{fig:intro} shows that collaborative signals are largely absent from these semantic graphs. Hence, enriching these semantic graphs with collaborative signals is necessary.

On the other hand, raw modality features are often corrupted by low-quality modality data, inducing structural noise in the constructed item-item semantic graphs and subsequently degrading the learned semantic representations. In the visual modality, background regions may dominate item images, hindering the extraction of item-specific features and resulting in spuriously high similarity between unrelated items. For example, a red ping pong paddle and a red dinner plate, both shown against white backgrounds, may be incorrectly perceived as visually similar. Likewise, in the textual modality, generic or promotional phrases such as ``high quality'' or ``best choice'' may overshadow key item-specific details, causing unrelated items to appear similar. Thus, it is essential to mitigate modality-induced noise to obtain high-quality semantic representations for recommendation.

% On the other hand, raw modality features of items often carry noise introduced by low-quality modality data, leading to structural distortion in the item-item semantic graphs constructed from them. For example, item images may contain distracting backgrounds that interfere with the accurate extraction of item-specific features, while textual descriptions can be filled with irrelevant content such as generic phrases or marketing terms (e.g., ``Top Quality'' or ``Best Choice'') \cite{xv2024improving}. As a result, the semantic representations derived from these graphs may be degraded, adversely affecting user preference modeling. Thus, it is essential to mitigate modality-induced noise to obtain high-quality representations for recommendation.

To address the aforementioned challenges, we propose \textbf{S}emantic \textbf{I}tem \textbf{G}raph \textbf{E}nhancement for multimodal \textbf{R}ecommendation (\textbf{SIGER}). Specifically, \textbf{SIGER} extracts item-item collaborative relations from user-item interaction data and integrates them into the modality-specific item-item semantic graphs to obtain modality-specific Enhanced Item-Item Semantic Graphs (EISG), thereby enabling comprehensive modeling of modality-aware and collaborative-aware signals. To mitigate the impact of structural noise in EISG caused by raw modality features, we propose a Modulus-based Personalized Embedding Perturbation mechanism. It injects both random and structure-aware perturbations into item representations to create contrastive views, with the perturbation intensity modulated in a personalized manner by a modulus-based strategy. Unlike uniform perturbation schemes, the personalized weighting mechanism ensures that the perturbation intensity remains appropriate for each item, avoiding both over-perturbation and under-perturbation. By extracting invariant information through contrastive learning, the model learns robust representations against modality noise. Finally, we propose a Dual Representation Alignment mechanism to better align behavior and semantic representations. It first employs a designed Anchor-based InfoNCE to align modality-specific semantic representations, using behavior representations as anchors, and then applies a standard InfoNCE to align the behavior representations with the fused semantic representations.
The main contributions are summarized as follows:
\begin{itemize}[leftmargin=*]
\vspace{-0.1em}
    \item We propose injecting collaborative signals into modality-specific item-item semantic graphs to enable unified modeling of modality-aware and collaboration-aware semantics, capturing more comprehensive item semantics.
    \vspace{-0.1em}
    \item We propose a Modulus-based Personalized Embedding Perturbation mechanism to mitigate structural noise in item semantic graphs. To our knowledge, this is the first to apply embedding perturbation for modality noise mitigation, and also the first to propose a modulus-based personalized strategy that adapts perturbation intensity suitably.
\vspace{-0.8em}
    \item We propose a Dual Representation Alignment mechanism that first uses the designed Anchor-based InfoNCE to align modality-specific semantics using behavior representations as anchors, and then uses a standard InfoNCE to align behavior and fused semantic representations.
    \vspace{-0.1em}
    \item We conduct extensive experiments on four datasets to assess the effectiveness of the proposed \textbf{SIGER} model.
\end{itemize}
\vspace{-0.5em}

\section{Related Work}
Early MMR methods usually integrate multimodal information into matrix factorization or collaborative filtering frameworks to improve user preference modeling \cite{xv2023commerce,chen2021graph,yang2021enhanced,jin2020multi}. A representative method, VBPR \cite{vbpr} incorporates visual features extracted by pre-trained CNN into item embeddings within a matrix factorization framework. 

% However, the direct incorporation of modality features may lead to interference from content unrelated to user preferences. DVBPR \cite{dvbpr} mitigates this issue by jointly learning visual representations and model parameters. VECF \cite{vecf} further introduces an attention-based framework that accounts for the non-uniform contribution of image regions to user preferences, achieving fine-grained visual preference modeling.

% Building on this idea, FREEDOM \cite{freedom} further filters out irrelevant connections through a pruning strategy that considers node degrees.

% MGCN \cite{mgcn} exploits behavior representations to infer user-specific preferences across modalities, which are then used to guide the adaptive integration of modality representations. 
\begin{figure*}[bpt]
    \centering
\includegraphics[width=0.99\textwidth]
{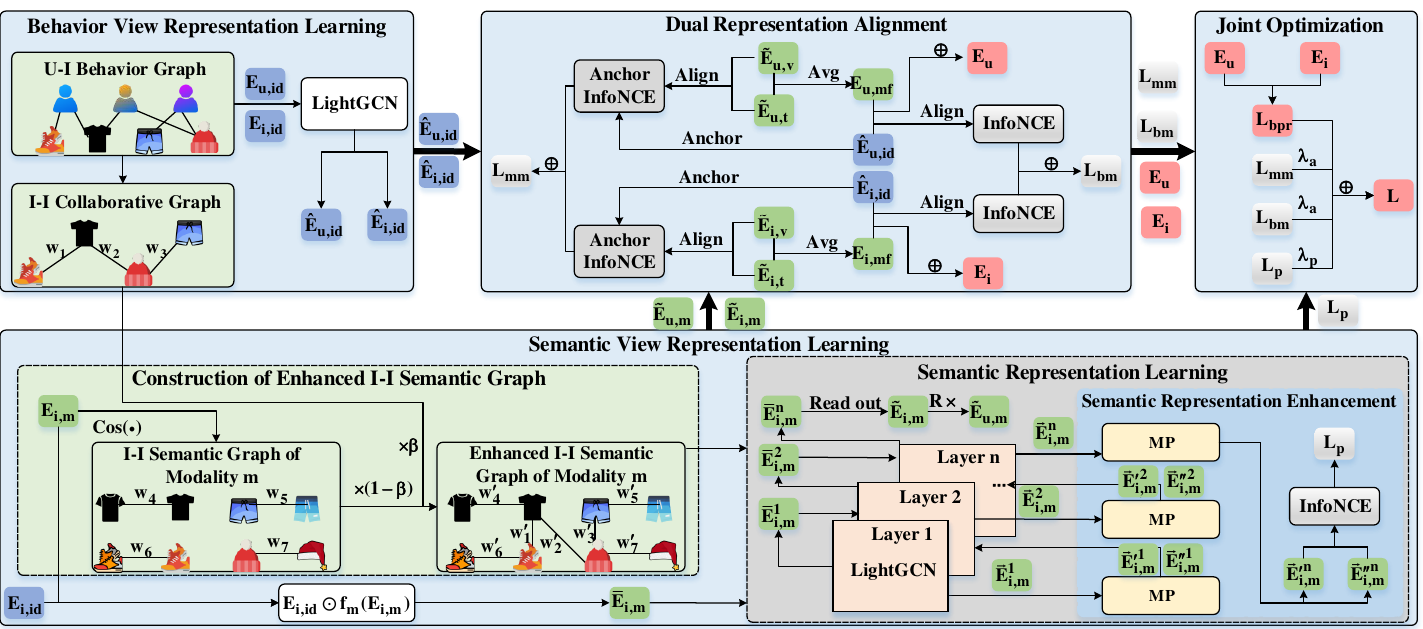} 
    
    \caption{The overview of \textbf{SIGER}. \textbf{First}, it encodes behavior representations with LightGCN. \textbf{Then}, it builds an enhanced item-item semantic graph by fusing modality- and collaborative-aware signals, and learns semantic representations with a noise-robust MP module. \textbf{Next}, it aligns behavior and semantic representations via Dual Representation Alignment. \textbf{Finally}, it jointly optimizes the BPR and auxiliary losses. Here, $m\in\{v,t\}$ denotes the visual and textual modality, while $u$ and $i$ denote any user and item, respectively. The subscript “$mf$” denotes the fused semantic representations, and $\textbf{E}$ is the embedding matrix.} 
    \label{fig:isge}  
    \vspace{-15pt}
\end{figure*}

Driven by the strength of Graph Convolutional Networks (GCNs) in capturing user preferences in conventional recommendation tasks~\cite{lightgcn,SelfCF2023TORS,LayerGCN2023ICDE}, various MMR methods based on GCNs have emerged~\cite{ mmsurvey2023arxiv}. MMGCN \cite{mmgcn} uses modality features as node attributes and leverages GCN propagation on the user–item graph to inject structural context into each modality, which are then fused into a unified representation for recommendation. Nevertheless, it fails to account for the structural noise inherent in the user–item graph. To address it, GRCN \cite{grcn} enhances the message passing mechanism in GCNs by adaptively weighting user–item interaction edges using multimodal signals. By assigning equal importance to all modalities, such methods neglect the modality-specific variations in user interests. DualGNN \cite{dualgcn} constructs a user–user co-occurrence graph and propagates weighted multimodal embeddings over it, enabling users to collaboratively learn personalized modality fusion patterns. Instead of propagating modality features on user–item graphs, LATTICE \cite{lattice} and MICRO \cite{micro} construct supplementary item–item graphs based on modality similarity and leverage them to refine modality representations, which are subsequently fused with behavioral signals for final recommendation. Despite their effectiveness, these approaches incur substantial computational overhead due to the requirement of updating modality-specific item–item semantic graphs during each training iteration. FREEDOM \cite{freedom} further reveals that that precomputing and freezing the graph structure prior to training provides a more efficient alternative without notable performance degradation, which is adopted in many subsequent studies \cite{mgcn, xv2024improving, dgvae2024tmm}. BM3 \cite{bm3} argues that auxiliary item–item graphs incur substantial computational and memory costs. It, instead, adopts a self-supervised learning framework that generates diverse contrastive views via embedding dropout and performs multi-level contrastive alignment to enhance representation quality. LGMRec \cite{lgmrec} point out that the above studies primarily focus on local user interests derived from interaction graphs, overlooking users’ global preferences—that is, their inherent interests in specific item attributes in modality data. It introduces a global hypergraph embedding module that captures global dependency relations to enhance preference modeling. MIG-GT \cite{hu2025modality} further finds that fixed GNN receptive fields (i.e., a fixed number of hops), commonly used in prior works, are suboptimal across modalities, and thus proposes modality-specific receptive fields to enhance performance. 
% CM$^3$ \cite{cm32025arxiv} revisits contrastive learning in multimodal models via the lens of alignment and uniformity. It proposes a calibrated uniformity loss to optimize the embedding space, ensuring semantically preserved item representations and achieving superior recommendation performance.
DA-MRS \cite{xv2024improving} simultaneously mitigates noise in user behaviors and multimodal content. It evaluates interaction confidence by multimodal signals to adapt the BPR loss for behavior denoising. Meanwhile, it improves each modality-specific item-item semantic graph through “consistent cross-modal similarities”, and enhances item representations by fusing representations from these refined graphs and an item-item behavior graph. However, embeddings from the behavior and modality item-item graphs are learned separately and averaged to produce the final representation. We argue that simple average ``post-fusion" may miss the complex interplay between modality-aware and interaction-aware relations. This motivates the construction of a unified item semantic graph to more effectively capture item-level semantics.  

\section{Preliminaries}
Let \( U = \{u\} \) and \( I = \{i\} \) be the user and item sets. The user-item interaction matrix is \(\mathbf{R} \in \{0,1\}^{|U| \times |I|}\), where \(\mathbf{R}_{u,i} = 1\) if user \(u\) interacted with item \(i\), else 0. 
User and item ID embeddings are \(\mathbf{E}_{id} \in \mathbb{R}^{(|U| + |I|) \times d}\), where \(d\) is the embedding dimension. Item modality features are \(\mathbf{E}_m \in \mathbb{R}^{|I| \times d_m}\), where \(d_m\) is modality-specific dimension, and \(m \in M = \{v, t\}\) denotes the \textbf{v}isual or \textbf{t}extual modality.

% The embedding of item \(i\) of modality \(m\) is denoted by \(\mathbf{E}_{i,m}\).
\section{Methodology}
In this section, we introduce the proposed \textbf{SIGER} by first presenting its overall architecture in Figure~\ref{fig:isge}, and then providing a detailed explanation.

\subsection{Behavior View Representation Learning}
This component captures interaction signals from the user-item graph by leveraging the GCN module in LightGCN \cite{lightgcn}. The update rule at the $l$-th convolution layer is: 
\begin{equation}
     \mathbf{E}^l_{id} = (\mathbf{D}^{-\frac{1}{2}}_\mathbf{G} \mathbf{G}\mathbf{D}^{-\frac{1}{2}}_\mathbf{G}) \mathbf{E}^{l-1}_{id},
\end{equation}
where \(\mathbf{E}_{id}^l\) is the ID embedding matrix at the \(l\)-th layer, 
\(\mathbf{D}_{\mathbf{G}}\) is the diagonal degree matrix of \(\mathbf{G}\) and \(\mathbf{G}\) is the symmetric matrix obtained by extending the interaction matrix \(\mathbf{R}\):
\[
     \mathbf{G} = 
 \begin{pmatrix}
 0 & \mathbf{R}\\
 \mathbf{R}^\top & 0
 \end{pmatrix}.
\]

The final ID embedding is derived by averaging all layers:
\begin{equation}
    \hat{\mathbf{E}}_{id} = \frac{1}{L+1}\sum^L_{l=0} \mathbf{E}^l_{id}.
\end{equation}

\subsection{Semantic View Representation Learning}
To strengthen semantics, we build an \textbf{E}nhanced \textbf{I}tem-Item \textbf{S}emantic \textbf{G}raph (\textbf{EISG}) and apply an \textbf{M}odulus-based Personalized Embedding \textbf{P}erturbation mechanism (\textbf{MP}) to lessen noise effects caused by raw modality features.

\noindent \textbf{\textit{Enhanced Item-Item Semantic Graph.}}
We first derive the item-item collaborative matrix $\mathbf{C}$ from user-item graph. Following \cite{he2021purify,xv2024improving,wang2020next,liang2016factorization}, the collaborative weight $\textbf{CR}$ for each item pair $(i, j)$ is computed using their co-occurrence frequency: 
\begin{equation}
    \text{CR}_{i,j} = Sigmoid(log~C_{i,j}),
\end{equation}
where $C_{i,j}$ is the co-occurrence frequency of $(i, j)$ in users' interaction list; Sigmoid function limits the weight to $[0, 1]$.

A larger CR means a stronger collaboration relation. To emphasize the most relevant neighbors, we retain only the Top-K neighbors with the highest CR values per item, following the Top-K strategy in \cite{lattice,chen2009fast, zhang2025communication}. Accordingly, the weight $\text{CR}_{i,j}$ is updated as: 
\begin{equation}\label{cr}
    \text{CR}_{i,j} = \left\{
\begin{array}{ll}
\text{CR}_{i,j}, & \text{CR}_{i,j} \in \text{Top-K}({\text{CR}_{i,k}~|~k \in I-i}). \\
0, & \text{Otherwise. }
\end{array}
\right.
\end{equation}

All $\text{CR}_{i,j}$ form the item-item collaborative matrix $\mathbf{C}$, which we further normalize to avoid gradient explosion:
\begin{equation}
    \overline{\mathbf{C}} = \mathbf{D}^{-\frac{1}{2}}_\mathbf{C} \mathbf{C} \mathbf{D}^{-\frac{1}{2}}_\mathbf{C},
\end{equation}
where $\mathbf{D}_\mathbf{C}$ denotes the diagonal degree matrix of $\mathbf{C}$. 

Next, we create the modality-specific item-item semantic graph \(\mathbf{H}_m\) for each modality \(m\). The semantic relation score \(\mathbf{H}^m_{i,j}\) between item \(i\) and \(j\) is calculated by computing the cosine similarity of their raw modality embeddings \(\mathbf{E}_{i,m}\) and \(\mathbf{E}_{j,m}\), then normalized to the range \([0,1]\). Formally:

\begin{equation}
    \text{H}^m_{i,j} = \frac{1+ s(\mathbf{E}_{i,m}, \mathbf{E}_{j,m})}{2}.
\end{equation}

\noindent where $s(\cdot)$ denotes cosine similarity.

Similar to the construction of \(\overline{\mathbf{C}}\), we only retain the Top-K highest-scoring neighbors (including the item itself, here) for each item in \(\mathbf{H}_m\), then apply normalization to get \(\overline{\mathbf{H}}_m\).

We then merge \(\overline{\mathbf{C}}\) with each \(\overline{\mathbf{H}}_m\) to form the modality-specific Enhanced Item-Item Semantic Graph \(\overline{\mathbf{S}}_m\): 

\begin{equation}\label{lc}
    \overline{\textbf{S}}_m = \beta \times \overline{\textbf{C}} + (1-\beta) \times \overline{\textbf{H}}_m,
\end{equation}
where $\beta$ is a shared hyper-parameter across all modalities. 

\noindent \textbf{\textit{Semantic Representation Learning.}} We then encode semantics of EISG. We use fused behavior and modality features as initial inputs, aligning with the fact that EISG merges collaborative and modality-aware semantic relation.

Due to dimension gap, we first map raw modality features to ID embedding space via a transformation function, and then fuse them by dot product. Formally: 
\begin{equation}\label{purify}
    \overline{\mathbf{E}}_{i,m} = \mathbf{E}_{i, id} \odot f_m(\mathbf{E}_{i,m}),
\end{equation}
where
\vspace{-0.5em}
\begin{equation}\label{fm}
    f_m(\mathbf{E}) = \sigma(\mathbf{W}''_m(\mathbf{W}'_m\mathbf{E}^\top + \mathbf{b}'_m)+\mathbf{b}''_m),
\end{equation}
and \(\mathbf{W}'_m \in \mathbb{R}^{d\times d_m}\), \(\mathbf{W}''_m \in \mathbb{R}^{d\times d}\), \(\mathbf{b}'_m, \mathbf{b}''_m \in \mathbb{R}^{d}\) are modality-specific parameters; ``\(\top\)'' denotes transposition.

Then, the \(l\)-th layer embedding update on \(\overline{\mathbf{S}}_m\) is:
\vspace{-0.5em}
\begin{equation}
    \overline{\mathbf{E}}^l_{i,m} = \overline{\mathbf{S}}_m \overline{\mathbf{E}}^{l-1}_{i,m}~,
\end{equation}
where $\overline{\mathbf{E}}_{i,m}$ in Eq. \eqref{purify} is used as the 0-th layer embeddings.

The final-layer embedding, denoted as \(\Tilde{\mathbf{E}}_{i,m}\), serves as the item semantics, while user semantics are derived by aggregating embeddings of interacted items:
\vspace{-0.5em}
\begin{equation}\label{ueu}
    \Tilde{\mathbf{E}}_{u,m} = (\mathbf{D}^{-\frac{1}{2}}_\mathbf{R} \mathbf{R}\mathbf{D}^{-\frac{1}{2}}_\mathbf{R}) \Tilde{\mathbf{E}}_{i,m}~,
\end{equation}
where $\mathbf{D}_\mathbf{R}$ is diagonal degree matrix of $\mathbf{R}$. 

We concatenate user and item modality-specific embeddings \(\Tilde{\mathbf{E}}_{u,m}\) and \(\Tilde{\mathbf{E}}_{i,m}\) as \(\Tilde{\mathbf{E}}_m \in \mathbb{R}^{(|U| + |I|) \times d}\).

\noindent \textbf{\textit{Semantic Representation Enhancement.}} To mitigate structural noise in EISG and enhance semantic encoding, we further propose a Modulus-based Personalized Embedding Perturbation mechanism to generate contrastive views for learning noise-robust representations. It performs additional graph convolutions on EISG using the same input $\overline{\mathbf{E}}_{i,m}$, while injecting personalized perturbations into each layer’s output \(\vec{\mathbf{E}}^{l}_{i,m}\). Specifically, normalized uniform noise is injected into \(\vec{\mathbf{E}}^{l}_{i,m}\) to simulate an embedding-level attack. The perturbed embeddings are then shuffled, re-normalized, and added back to the originals to mimic a structure-level disturbance. Moreover, a modulus-based perturbation weighting mechanism is used to personalize perturbation intensity across items based on their embedding modulus. Formally, the perturbed embedding for the $l$-th embedding $\vec{\mathbf{E}}^{l}_{i,m}$, denoted by ${\vec{\mathbf{E}}}^{'l}_{i,m}$, is obtained as follows:
\vspace{-0.3em}
\begin{equation}
\begin{aligned}
\mathbf{W}_p ~= &~~ \frac{||\vec{\mathbf{E}}^{l}_{i,m}||_{row,2}}{max(||\vec{\mathbf{E}}^{l}_{i,m}||_{row,2})};\\
   \dot{\mathbf{E}}^{l}_{i,m} = & ~~\vec{\mathbf{E}}^{l}_{i,m} + \epsilon \times \mathbf{W}_p \times sign(\vec{\mathbf{E}}^{l}_{i,m}) \odot f_{norm}(\Delta);\\
    {\vec{\mathbf{E}}}^{'l}_{i,m} = & ~~ \vec{\mathbf{E}}^{l}_{i,m} + \epsilon \times \mathbf{W}_p \times f_{shuffle}(f_{norm}(\dot{\mathbf{E}}^{l}_{i,m})),
\end{aligned}
\end{equation}

\noindent where $||\cdot||_{row,2}$ is row-wise Frobenius norm; \(\mathbf{W}_p \in \mathbb{R}^{|I|}\) denotes personalized perturbation weights; \(\Delta \sim U(0,1) \in \mathbb{R}^{|I| \times d}\); \(\text{sign}(\cdot)\) aligns perturbation direction with the embedding; \(f_{\text{norm}}(\cdot)\) normalizes row vectors; \(f_{\text{shuffle}}(\cdot)\) randomly shuffles rows; and \(\epsilon\) controls global perturbation intensity.

We simplify the final-layer perturbed embedding \({\vec{\mathbf{E}}}^{'n}_{i,m}\) as \(\Tilde{\mathbf{E}}'_{i,m}\). A second view \(\Tilde{\mathbf{E}}''_{i,m}\) is obtained by repeating the process. We enhance representations by maximizing the InfoNCE-based mutual information between the two views:
\vspace{-0.5em}
\begin{equation}
    \mathcal{L}_p = \sum_{m \in M}\frac{1}{|I|}\sum_{i \in I} -log \frac{exp(s(\Tilde{\mathbf{E}}'_{i,m}, \Tilde{\mathbf{E}}''_{i,m})/\tau_0)}{\sum_{i' \in I} exp(s(\Tilde{\mathbf{E}}'_{i,m}, \Tilde{\mathbf{E}}''_{i',m})/\tau_0)},
\end{equation}
where $s(\cdot)$ denotes cosine similarity; $\tau_0$ is the temperature.

\subsection{Dual Representation Alignment}

To mitigate the disparity of representation spaces, we propose a new \textbf{D}ual Representation \textbf{A}lignment (\textbf{DA}) strategy to align multiple representations. It first aligns different semantic representations and then aligns the fused semantics with behavior representations.

\noindent \textbf{\textit{Semantic Representations Alignment.}} We propose Anchor-based InfoNCE, which uses behavior representations as anchors to align multiple modality-specific semantic representations. Formally, item-level alignment is:
\vspace{-0.25em}
\begin{equation}
\begin{aligned}
    \mathcal{L}^I_{mm} = &\frac{1 }{|I|} \sum_{i\in I} -log( \frac{A_{i, id}}{A_{i, id}  + \sum_{i'\in I-i} exp(s(\Tilde{\mathbf{E}}_{i,v}, \Tilde{\mathbf{E}}_{i',t})/\tau_1)});\\
    \\
    A_{i, id} = &~ exp( \frac{s(\Tilde{\mathbf{E}}_{i,v}, \hat{\mathbf{E}}_{i,id}) + s(\Tilde{\mathbf{E}}_{i,t}, \hat{\mathbf{E}}_{i,id})}{2\tau_1}) ,
\end{aligned}
\end{equation}
where $s(\cdot)$ denotes cosine similarity; $\tau_1$ is the temperature.

Similarly, the user-side loss is \(\mathcal{L}^U_{mm}\). The total semantic representation alignment loss is $\mathcal{L}_{mm} = \mathcal{L}^U_{mm} + \mathcal{L}^I_{mm}$.

\noindent \textbf{\textit{Behavior and Semantic Representations Alignment.}} Aligned semantic representations are first fused by average:
\begin{equation}
    \mathbf{E}_{mf} = \frac{1}{|M|} \sum_{m\in M} \Tilde{\mathbf{E}}_{m}.
\end{equation}
The fused semantic representations are further aligned with behavior representations:
\begin{equation}
\begin{aligned}
    \mathcal{L}_{bm} = & \frac{1}{|U|}\sum_{u\in U} -log\frac{ exp(s(\mathbf{E}_{u,mf}, \hat{\mathbf{E}}_{u,id}) / \tau_2))}{\sum_{u'\in U} exp(s(\mathbf{E}_{u,mf}, \hat{\mathbf{E}}_{u',id}) / \tau_2))}\\  
    + &\frac{1}{|I|}\sum_{i\in I} -log\frac{ exp(s(\mathbf{E}_{i,mf}, \hat{\mathbf{E}}_{i,id}) / \tau_2))}{\sum_{i'\in I} exp(s(\mathbf{E}_{i,mf}, \hat{\mathbf{E}}_{i',id}) / \tau_2))}.
\end{aligned}
\end{equation}

Final user and item representations are formed by merging behavior and fused semantic representations:
\begin{equation}
\mathbf{E} = \hat{\mathbf{E}}_{id} + \mathbf{E}_{mf}.
\end{equation}

\subsection{Model Optimization}
We define interaction probability of user-item pair $(u, i)$ as the inner product of user and item representations:
\begin{equation}
    \hat{y}_{ui} = \mathbf{E}_{u} \cdot  \mathbf{E}_{i}.
\end{equation}

\noindent Then, we adopt BPR \cite{lattice} as the main objective function, yielding loss \(\mathcal{L}_{bpr}\). Next, combining it with alignment and perturbation losses to jointly optimize the representations:
\vspace{-0.3em}
\begin{equation}
    \mathcal{L} = \mathcal{L}_{bpr} + \lambda_p \mathcal{L}_p + \lambda_{a}(\mathcal{L}_{mm} +  \mathcal{L}_{bm}) +\lambda_r||\mathbf{E}||_2,
\end{equation}
where $\lambda_p$, $\lambda_{a}$ and $\lambda_{r}$ are constant parameters; $||\mathbf{E}||_2$ is the $L_2$ regularization term.

\section{Experiments}
We assess \textbf{SIGER} on four datasets to answer the questions:

\noindent \textbf{RQ1:} To what extent does \textbf{SIGER}, exceed SOTA models?

\noindent \textbf{RQ2:} To what extent can \textbf{SIGER} mitigate cold-start issue?

\noindent \textbf{RQ3:} What is the impact of each key component of \textbf{SIGER} on overall recommendation performance? 

\noindent \textbf{RQ4:} What is the impact of critical hyperparameters on the performance of \textbf{SIGER}?
% \noindent \textbf{RQ4:} What is the impact of neighbor size in EISG's collaborative graph on performance? 

% \noindent \textbf{RQ5:} What role does the number of convolution layers of EISG play in the performance of \textbf{SIGER}?

\begin{table*}[t]
    \def\arraystretch{1.25}	
    \centering
    \small
    \tabcolsep=0.03cm
    \renewcommand{\arraystretch}{0.0001}
    \begin{tabular}{lcccccccccccccc}
        \toprule
        \multirow{2}{*}{Datasets}& \multirow{2}{*}{Metrics} & \fontsize{8.5}{9.5}\selectfont \textbf{LightGCN} && \fontsize{8.5}{9.5}\selectfont \textbf{VBPR} & \fontsize{8.5}{9.5}\selectfont \textbf{LATTICE} & \fontsize{8.5}{9.5}\selectfont \textbf{SLMRec} &\fontsize{8.5}{9.5}\selectfont \textbf{FREEDOM} &\fontsize{8.5}{9.5}\selectfont \textbf{BM3} & \fontsize{8.5}{9.5}\selectfont \textbf{MGCN} & \fontsize{8.5}{9.5}\selectfont \textbf{LGMRec} & \fontsize{8.5}{9.5}\selectfont \textbf{DA-MRS} & \fontsize{8.5}{9.5}\selectfont \textbf{MIG-GT} & \fontsize{8.5}{9.5}\selectfont \textbf{SIGER} & \fontsize{8.5}{9.5}\selectfont \textbf{Impro.}\\ [ 0.2 em]
        % \cline{3-3}
        % \cline{5-14}
                                                             & &\fontsize{8.5}{9.5}\selectfont SIGIR'20 && \fontsize{8.5}{9.5}\selectfont AAAI'16 & \fontsize{8.5}{9.5}\selectfont MM'21 &\fontsize{8.5}{9.5}\selectfont  TMM'22 & \fontsize{8.5}{9.5}\selectfont MM'23 &  \fontsize{8.5}{9.5}\selectfont WWW'23 & \fontsize{8.5}{9.5}\selectfont  MM'23 &\fontsize{8.5}{9.5}\selectfont AAAI'24 &\fontsize{8.5}{9.5}\selectfont  KDD'24 &\fontsize{8.5}{9.5}\selectfont AAAI'25 & \fontsize{8.5}{9.5}\selectfont Ours & \\
        \midrule
        \multirow{4}{*}{Baby}     &R@10 &0.0479 &&0.0423 &0.0547 &0.0547 &0.0624 &0.0542 &0.0607 &0.0645 &0.0650 &\underline{0.0665}   &\textbf{0.0697} & 4.81\%\\
                                  &R@20 &0.0754 &&0.0664 &0.0844 &0.0810 &0.0985 &0.0862 &0.0950 &0.0981 &0.0994 &\underline{0.1021}   &\textbf{0.1060} & 3.82\%\\
                                  &N@10   &0.0257 &&0.0223 &0.0289 &0.0285 &0.0324 &0.0285 &0.0328 &0.0350 &0.0346 &\underline{0.0361}   &\textbf{0.0376} &4.16\%\\
                                  &N@20   &0.0328 &&0.0284 &0.0366 &0.0357 &0.0416 &0.0367 &0.0416 &0.0437 &0.0435 &\underline{0.0452}   &\textbf{0.0469} &3.76\%\\
        \midrule

        \multirow{4}{*}{Sports}   &R@10 &0.0569 &&0.0561 &0.0626 &0.0676 &0.0713 &0.0619 &0.0737 &0.0724 &0.0751 &\underline{0.0753}   &\textbf{0.0810} &7.57\%\\
                                  &R@20 &0.0864 &&0.0859 &0.0959 &0.1017 &0.1077 &0.0971 &0.1107 &0.1087 &0.1125 &\underline{0.1130}   &\textbf{0.1197} &5.93\%\\
                                  &N@10   &0.0311 &&0.0307 &0.0337 &0.0374 &0.0382 &0.0338 &0.0403 &0.0392 &0.0402 &\underline{0.0414}   &\textbf{0.0441} &6.52\%\\
                                  &N@20   &0.0387 &&0.0384 &0.0423 &0.0462 &0.0476 &0.0429 &0.0499 &0.0485 &0.0498 &\underline{0.0511}   &\textbf{0.0541} &5.87\%\\
        \midrule

        \multirow{4}{*}{Clothing} &R@10 &0.0361 &&0.0283 &0.0468 &0.0540 &0.0624 &0.0425 &\underline{0.0658} &0.0549 &0.0647 &0.0636   &\textbf{0.0689} &4.71\%\\
                                  &R@20 &0.0544 &&0.0417 &0.0688 &0.0810 &0.0928 &0.0637 &\underline{0.0963} &0.0822 &\underline{0.0963} &0.0934   &\textbf{0.1000} &3.84\%\\
                                  &N@10   &0.0197 &&0.0157 &0.0256 &0.0285 &0.0336 &0.0232 &\underline{0.0359} &0.0301 &0.0353 &0.0347   &\textbf{0.0369} &2.79\%\\
                                  &N@20   &0.0243 &&0.0191 &0.0312 &0.0357 &0.0414 &0.0286 &\underline{0.0436} &0.0370 &0.0433 &0.0422   &\textbf{0.0450} &3.21\%\\

        \midrule

        \multirow{4}{*}{MicroLens} &R@10 &0.0720 &&0.0677 &0.0726 &0.0778 &0.0674 &0.0606 &0.0756 &0.0748 &\underline{0.0815} &0.0806   &\textbf{0.0875} &7.36\%\\
                                  &R@20 &0.1075 &&0.1026 &0.1089 &0.1190 &0.1032 &0.0981 &0.1134 &0.1132 &\underline{0.1221} &0.1189   &\textbf{0.1279} &4.75\%\\
                                  &N@10   &0.0376 &&0.0351 &0.0380 &0.0405 &0.0345 &0.0304 &0.0387 &0.0390&\underline{0.0431} &0.0426   &\textbf{0.0460} &6.73\%\\
                                  &N@20   &0.0467 &&0.0441 &0.0473 &0.0511 &0.0437 &0.0400 &0.0484 &0.0489 &\underline{0.0536} &0.0523   &\textbf{0.0564} &5.22\%\\
        
        \bottomrule
    \end{tabular}
    \caption{Performance comparison of recommendation models. The \textbf{bold} font highlights the best result, \underline{underline} indicates the second best, and Impro. denotes the relative improvement over the strongest baseline.}
    \label{tab:me}
    \vspace{-1em}
\end{table*}
\subsection{Experimental Configuration}
\noindent \textbf{Dataset.} We use four datasets: Baby, Sports, Clothing, and MicroLens \cite{ni2023content}. Modality features are extracted following the standard practice in \cite{freedom}.

\noindent \textbf{Baselines.} We adopt SOTA models from two categories: (1) model using only interaction data, represented by \textbf{LightGCN} \cite{lightgcn}; and (2) multimodal models, including \textbf{VBPR} \cite{vbpr}, \textbf{LATTICE} \cite{lattice}, \textbf{SLMRec} \cite{slmrec}, \textbf{FREEDOM} \cite{freedom}, \textbf{MGCN} \cite{mgcn}, \textbf{BM3} \cite{bm3}, \textbf{LGMRec} \cite{lgmrec}, \textbf{DA-MRS} \cite{xv2024improving} and \textbf{MIG-GT} \cite{hu2025modality}.

\noindent \textbf{Evaluation Protocols.}
To thoroughly evaluate model performance, we conduct both general and cold-start evaluations. For both cases, we use Recall@K (R@K) and NDCG@K (N@K) with K = 10 and 20 as metrics. The dataset is split following the widely adopted protocol in \cite{lattice}.
Training stops if R@20 does not improve for 20 consecutive epochs.

\noindent \textbf{Implementation Specifications.} We implement all models using the MMRec framework \cite{zhou2023mmrec} on Nvidia A100 with 40G. For baselines, we use the optimal hyperparameters reported in their original papers. For our model, the Top-K values for the modality-specific semantic graph and the collaborative relation graph are set to 10 and 5, respectively. The number of GCN layers for the user-item graph and EISG is set to 3 and 2, respectively. The learning rate is set to $1 \times 10^{-3}$. $\epsilon = 0.05$. $\lambda_r = 1\times 10^{-5}$. Remaining hyperparameters are tuned via grid search: $\beta \in \{0.1, 0.2, 0.3, 0.4, 0.5\}$, $\tau_{1/2} \in \{0.1, 0.2\}$ and $\lambda_{p/a} \in \{0.005, 0.01, 0.02, 0.03\}$. 

% The optimal setting of these fine-tuned parameters on four datasets is provided in the accompanying code. 

\subsection{Performance Comparison}
\noindent \textbf{General Evaluation. (RQ1)} Table \ref{tab:me} compares the performance between \textbf{SIGER} and baselines in the general setting.

From the results, we can find that: (1) The proposed \textbf{SIGER} achieves substantial gains over all baselines across all metrics. In particular, on the Sports and MicroLens datasets, it exceeds the strongest baseline by over 6.5\% on half of the metrics, with a peak improvement of 7.57\%. For most of the remaining metrics, the improvements surpass 5\%, and even the smallest gain remains close to this margin at 4.75\%. Despite being less pronounced than on Sports and MicroLens, the improvement of \textbf{SIGER} on the Baby and Clothing datasets remains noteworthy. Only one metric shows a gain slightly below 3\% (2.79\%), while nearly half of the metrics exceed 4\%, with the highest reaching 4.81\%. The results confirm that \textbf{SIGER} effectively models user preferences and yields superior recommendation performance.
% (2) Multimodal information indeed helps improve performance, as evidenced by the fact that all evaluated multimodal models, except for \textbf{VBPR}, are built upon the non-multimodal \textbf{LightGCN} and consistently surpass it in accuracy. 
(2) Effectively fusing behavior and semantic representations is essential. For instance, \textbf{LATTICE} obtains item semantics from modality-driven item-item graphs and merges them with behavior representations through simple summation. \textbf{MGCN} builds upon this by introducing InfoNCE-based contrastive learning to align the two before fusion. In contrast, our \textbf{SIGER} incorporates a novel Dual Representation Alignment strategy that first aligns multiple semantic views using behavior representations as anchors, followed by alignment between the fused semantic and behavior representations, leading to further improvements.

\begin{figure}[bpt]
    \centering
\includegraphics[trim={0 6 0 6},clip,width=0.48\textwidth]{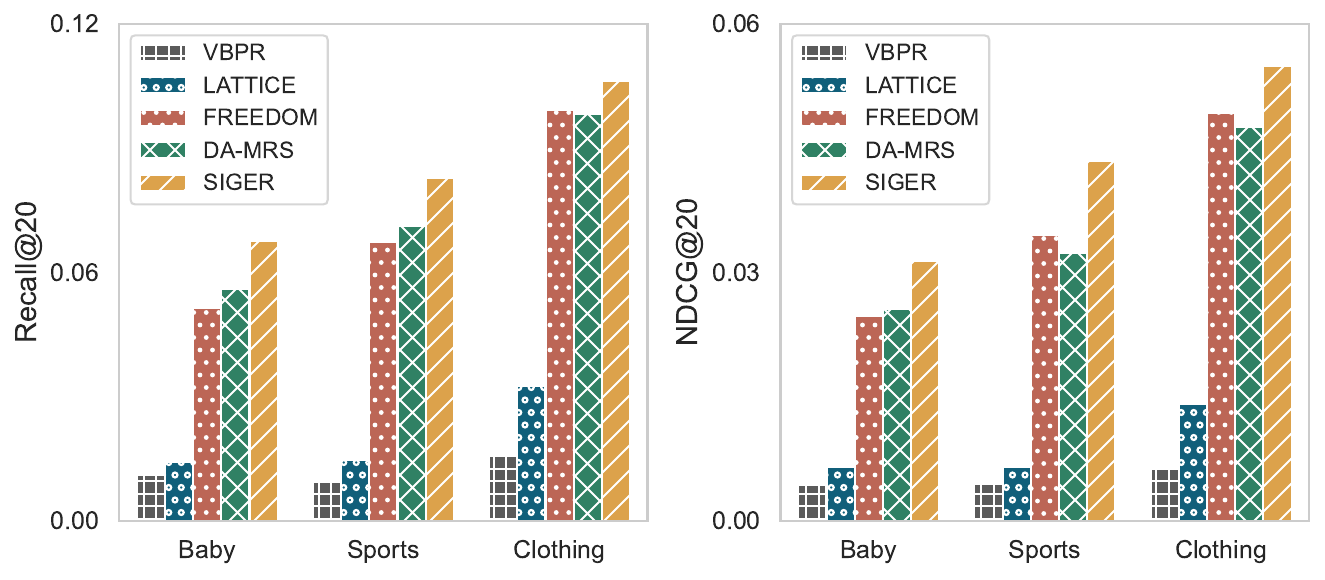} 
    
    \caption{Comparison of \textbf{SIGER} with several recent models in cold-start scenarios.}  
    \label{fig:coldstart}  
    \vspace{-2em}
\end{figure}

\noindent \textbf{Cold-Start Evaluation (RQ2).} 
We also assess \textbf{SIGER} in cold-start scenarios against some recent baselines, as shown in Figure~\ref{fig:coldstart}. The results show that \textbf{SIGER} can significantly improve performance in cold-start setting compared with the baselines.
It yields 20.32\%, 16.59\%, and 7.06\% relative gains in Recall@20 over the best-performing baseline on Baby, Sports, and Clothing, respectively. The performance gain can be attributed to two factors. First, EISG fuses collaborative and modality-based item semantics, enabling more effective message propagation and making cold-start items more ``visible''. Second, we reduce the impact of raw modality noise on EISG to ensure more accurate representations. These notable gains, together with those of general setting, validate \textbf{SIGER}’s effectiveness in diverse scenarios.

\subsection{Ablation Study (RQ3)} \label{abe}
% \vspace{-0.4em}
This section first provides a detailed examination of the contribution of each component to overall performance. And then examine the impact of different modality features.

\noindent  \textbf{1) \textit{Component-Related Ablation Study}}: We construct the following component-related variants: (1) \textbf{SIGER/EISG}, replacing EISG with an item-item semantic graph built only from raw modality features; (2) \textbf{SIGER/MP} and (3) \textbf{SIGER/DA}, removing the MP and DA components, respectively; (4) \textbf{SIGER/MP*}
, only removing the modulus-based perturbation weighting mechanism in MP;  (5) \textbf{SIGER/DA*}
, replacing the Anchor-based InfoNCE in DA with standard InfoNCE to test whether behavior representations as anchors improve semantic alignment. The latter two are fine-grained variants that further examine the subcomponents of MP and DA. Table \ref{tab:ae} presents the results. 

\begingroup
\begin{table}
    \centering
    \fontsize{5.5}{7.5}\selectfont
    \tabcolsep=0.03cm
    \renewcommand{\arraystretch}{0.6} 
    \begin{tabular}{llcccccc}
        \toprule
        Datasets& Metrics &  \fontsize{6}{9.5}\selectfont \textbf{SIGER} & \fontsize{6}{9.5}\selectfont \textbf{SIGER/EISG} & \fontsize{6}{9.5}\selectfont \textbf{SIGER/MP} & \fontsize{6}{9.5}\selectfont \textbf{SIGER/DA} & \fontsize{6}{9.5}\selectfont \textbf{SIGER-MP*} & \fontsize{6}{9.5}\selectfont \textbf{SIGER-DA*}\\
        \midrule
        \multirow{4}{*}{Baby}     &R@10 &\textbf{0.0697} &0.0639 &0.0673 &0.0662 & 0.0678 &0.0679 \\
                                  &R@20 &\textbf{0.1060} &0.0985 &0.1024 &0.1043 & 0.1053 &0.1032\\
                                  &N@10   &\textbf{0.0376} &0.0341 &0.0369 &0.0355 & 0.0366 &0.0366\\
                                  &N@20   &\textbf{0.0469} &0.0429 &0.0458 &0.0452 & 0.0463 &0.0457\\
        \midrule

        \multirow{4}{*}{Sports}   &R@10 &\textbf{0.0810} &0.0677 &0.0781 &0.0768 & 0.0790 &0.0775\\
                                  &R@20 &\textbf{0.1197} &0.1024 &0.1159 &0.1162 & 0.1171 &0.1169\\
                                  &N@10   &\textbf{0.0441} &0.0362 &0.0426 &0.0416 & 0.0428 &0.0424\\
                                  &N@20   &\textbf{0.0541} &0.0452 &0.0524 &0.0517 & 0.0526 &0.0525\\
        \midrule

        \multirow{4}{*}{Clothing} &R@10 &\textbf{0.0689} &0.0631 &0.0665 &0.0642 & 0.0646 &0.0673\\
                                  &R@20 &\textbf{0.1000} &0.0939 &0.0984 &0.0953 & 0.0971 &0.0969\\
                                  &N@10   &\textbf{0.0369} &0.0340 &0.0361 &0.0343 & 0.0352 &0.0364\\
                                  &N@20   &\textbf{0.0450} &0.0418 &0.0442 &0.0422 & 0.0435 &0.0439\\

        \midrule

        \multirow{4}{*}{MicroLens} &R@10 &\textbf{0.0875} &0.0722 &0.0867 &0.0837 & 0.0868 &0.0863\\
                                  &R@20 &\textbf{0.1279} &0.1083 &0.1272 &0.1244 & 0.1278 &0.1255\\
                                  &N@10   &0.0460 &0.0372 &\textbf{0.0461} &0.0439 & 0.0457 &0.0460\\
                                  &N@20   &0.0564 &0.0465 &\textbf{0.0566} &0.0544 & 0.0562 &0.0562\\
        \bottomrule
    \end{tabular}
        \caption{Performance comparison between SIGER with its component-related variants.}
        \label{tab:ae}
    \vspace{-20pt}
\end{table}
\endgroup

First, by comparing SIGER with its coarse-grained variants (SIGER/EISG, SIGER/MP, and SIGER/DA), we find that each component contributes to overall performance, confirming their effectiveness. Among them, EISG contributes the most, DA comes next, and MP has the least effect overall.  Despite MP's relatively weak contributions, it still yields relative improvements in R@10 of 3.57\%, 1.41\%, 3.01\% on the Baby, Sports, and Clothing datasets, respectively. On MicroLens, MP yields only marginal or no improvement, which is likely because the raw modality features contain less noise, limiting the benefit of perturbation.

% For instance, on the Clothing dataset, the three components yield relative improvements in R@20 of 6.1\%, 4.7\%, and 1.6\%, respectively.

Second, the degradation of SIGER-MP* compared to SIGER shows that the modulus-based perturbation weighting mechanism in MP plays a crucial role in boosting performance. The decline of SIGER-MP* relative to SIGER/MP on Clothing further supports this, showing that applying embedding perturbation with uniform rather than modulus-guided personalized intensity may harm performance. This necessity of adapting perturbation intensity across items stems from discrepancies in item embedding modulus, where applying same perturbations may lead to inconsistent perturbation intensities across items. For some, this may introduce overwhelming noise and degrade performance. Adjusting perturbation intensity according to each item's embedding modulus improves perturbation robustness.

Third, SIGER’s superior performance over SIGER-DA* shows benefit of DA's Anchor-based InfoNCE sub-component, which uses behavior representations to guide semantic representation alignment. This gain may be attributed to its ability to mitigate the dominance of one semantic representation during alignment, where one representation is overly optimized toward the other while the latter remains nearly unchanged. By treating the behavior representations as anchors, the alignment process leads both semantic representations to adjust toward it more effectively.

\noindent  \textbf{2) \textit{Modality-Related Ablation Study}}: We remove the visual and textual modality features to obtain the variants \textbf{SIGER/V} and \textbf{SIGER/T}, respectively. Since each variant retains only one modality, we remove the semantic representation alignment and retain only the behavior-semantic alignment in DA. The results are shown in Table~\ref{tab:mae}.
\begingroup
\begin{table}[t]
    \centering
    \fontsize{7.5}{9}\selectfont
    \def\arraystretch{0.8}
    \tabcolsep=0.28cm
    \begin{tabular}{llcccc}
        \toprule
        Dataset & Model & R@10 & R@20 & N@10 & N@20 \\
        \midrule
        \multirow{3}{*}{Baby} 
            & SIGER/T  & 0.0603 & 0.0943 & 0.0331 & 0.0418 \\
            & SIGER/V  & 0.0678 & 0.1047 & 0.0367 & 0.0462 \\
            & SIGER    & \textbf{0.0697} & \textbf{0.1060} & \textbf{0.0376} & \textbf{0.0469} \\
        \midrule
        \multirow{3}{*}{Sports} 
            & SIGER/T  & 0.0713 & 0.1066 & 0.0386 & 0.0477 \\
            & SIGER/V  & 0.0779 & 0.1164 & 0.0427 & 0.0527 \\
            & SIGER    & \textbf{0.0810} & \textbf{0.1197} & \textbf{0.0441} & \textbf{0.0541} \\
        \midrule
        \multirow{3}{*}{Clothing} 
            & SIGER/T  & 0.0544 & 0.0792 & 0.0298 & 0.0361 \\
            & SIGER/V  & 0.0658 & 0.0958 & 0.0357 & 0.0433 \\
            & SIGER    & \textbf{0.0689} & \textbf{0.1000} & \textbf{0.0369} & \textbf{0.0450} \\
        \midrule
        \multirow{3}{*}{MicroLens} 
            & SIGER/T  & 0.0831 & 0.1207 & 0.0438 & 0.0535 \\
            & SIGER/V  & 0.0833 & 0.1220 & 0.0441 & 0.0541 \\
            & SIGER    & \textbf{0.0875} & \textbf{0.1279} & \textbf{0.0460} & \textbf{0.0564} \\
        \bottomrule
         
    \end{tabular}
    \caption{Performance comparison of SIGER and its modality-related variants.}
    \label{tab:mae}
    \vspace{-2em}
\end{table}
\endgroup

From Table~\ref{tab:mae}, both modalities clearly enhance performance, as \textbf{SIGER}/V and \textbf{SIGER}/T consistently underperform \textbf{SIGER}. Notably, \textbf{SIGER}/V outperforms \textbf{SIGER}/T across all datasets and metrics. For instance, it achieves relative R@20 gains of 11.03\%, 9.19\%, 20.96\%, and 1.08\% on Baby, Sports, Clothing, and MicroLens, respectively, highlighting the greater importance of textual information in modeling user preferences. Moreover, combining Tables~\ref{tab:me} and~\ref{tab:mae} shows that \textbf{SIGER/V}, despite using only text modality, outperforms all multimodal baselines on Baby and Sports, with R@20 gains of 5.3\% and 3.47\% over the strongest baseline, respectively. It also slightly surpasses the best baseline on MicroLens and performs comparably on Clothing, further validating the strength of our model. 
            
\subsection{Hyperparameter Analysis (\textbf{RQ4})}
\begin{figure}[h]
    \centering
\includegraphics[width=0.48\textwidth]{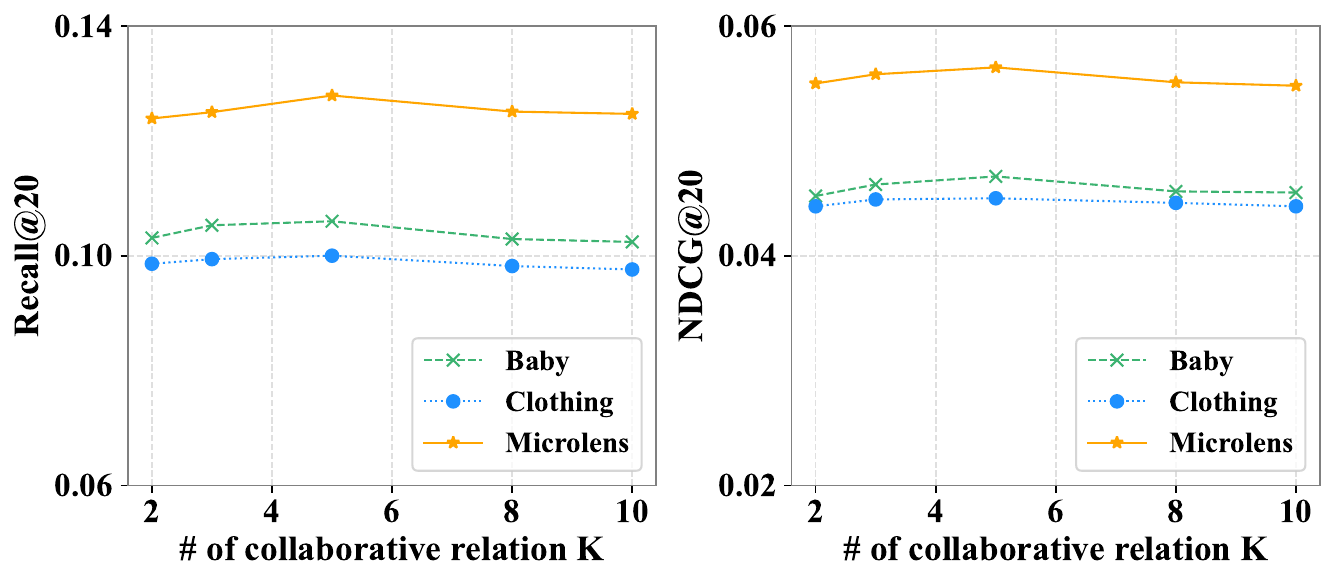} 
    \caption{Performance of \textbf{SIGER} with respect to varying numbers of collaborative relations $\text{K}$.}  
    \label{fig:para_k}  
    
\end{figure}

\begin{figure}[h]
    \centering
\includegraphics[width=0.48\textwidth]{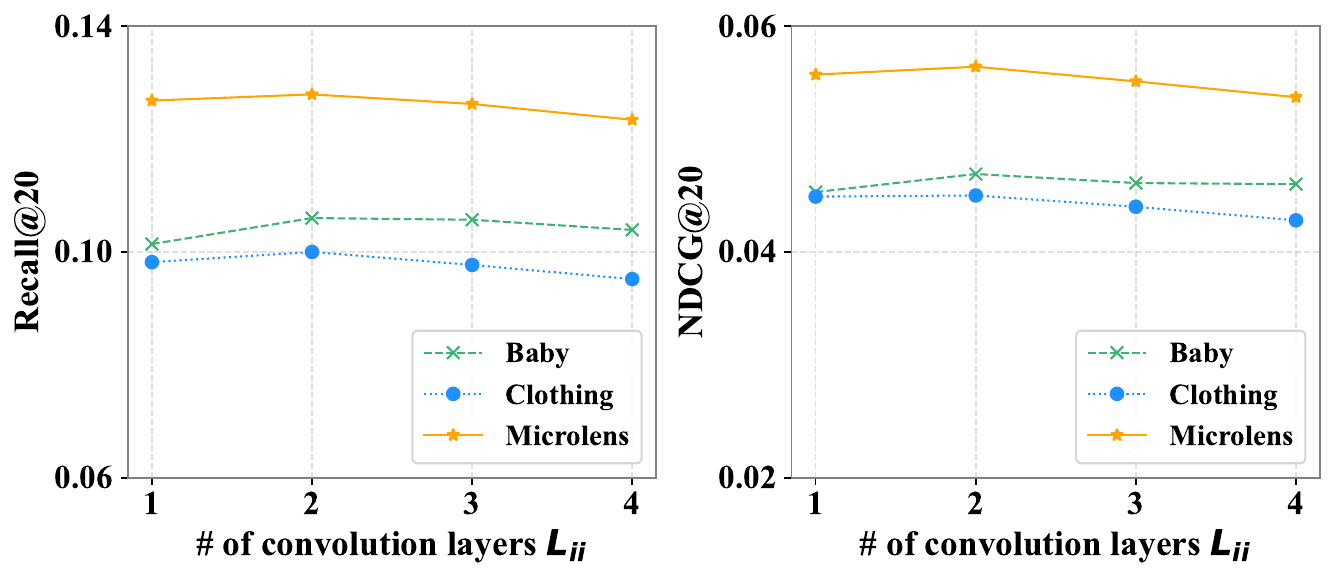} 
    \caption{Performance of \textbf{SIGER} with respect to varying convolution layers $L_{ii}$ on EISG.}  
    \label{fig:para_layer}  
    \vspace{-2em}
\end{figure}

This section investigates the impact of two essential hyperparameters in the EISG module on the performance of \textbf{SIGER}: 1) The $\text{K}$ in Equation \eqref{cr} (The number of selected collaborative neighbors). 2) $L_{ii}$ (The number of convolution layers applied to the EISG). The $\text{K}$ is selected from $\{2, 3, 5, 8, 10\}$ and $L_{ii}$ from $\{1, 2, 3, 4\}$. The corresponding results are shown in Figure~\ref{fig:para_k} and~\ref{fig:para_layer}, respectively.

Figure~\ref{fig:para_k} shows that as $\text{K}$ increases from 2 to 10, \textbf{SIGER}'s performance in terms of Recall@20 and NDCG@20 first improves and then drops, especially on Baby and Clothing. For instance, on the Baby dataset, Recall@20 peaks at 0.1060 when $\text{K} = 5$. As $\text{K}$ increases from 2 to 3, the metric rises from 0.1031 to 0.1053, but then declines from 0.1029 to 0.1024 as $\text{K}$ grows from 8 to 10. NDCG@20 witnesses a similar trend. A similar pattern is also observed on the MicroLens dataset. Clothing shows weaker sensitivity to the variation in collaborative neighbor counts within this range. The highest Recall@20 (0.1000) occurs at $\text{K}=5$, only 0.0006 higher than the second-highest value at $\text{K}=3$. Results for other $\text{K}$ values are slightly lower, fluctuating around 0.0985 and remaining close to the peak. The limited sensitivity to $\text{K}$ on the Clothing dataset is consistent with Table~\ref{tab:me}, where its gains are smaller than those on other datasets. Overall, $\text{K} = 5$ yields the best results on all datasets.

Figure~\ref{fig:para_layer} reveals a trend similar to Figure~\ref{fig:para_k}, where increasing the number of graph convolution layers $L_{ii}$ from 1 to 4 first enhances performance, followed by a decline. For example, on MicroLens, Recall@20 rises steadily from 0.1268 to 0.1279 as $L_{ii}$ increases from 1 to 2, but then drops from 0.1262 to 0.1234 as $L_{ii}$ goes from 3 to 4. The best performance is achieved when $L_{ii}$ is set to 2. NDCG@20 follows a similar trend. Similarly, the Baby and Clothing datasets show the same trend as MicroLens. In general, setting the number of graph convolution layers to 2 yields the optimal performance of \textbf{SIGER} across all datasets.

\section{Conclusion}

In this paper, we present SIGER, a framework that enhances user preference modeling by capturing more comprehensive and accurate item semantic relations. We first construct an Enhanced Item-Item Semantic Graph by integrating collaborative signals from the user-item graph with modality-based similarities derived from raw modality features, to learn enriched item semantics. To alleviate the impact of structural noise in semantic graph, we introduce a modulus-based personalized embedding perturbation strategy. This module perturbs embeddings with modulus-guided personalized intensity to produce contrastive views and employs contrastive learning to obtain noise-robust representations. Furthermore, we design a dual representation alignment mechanism to effectively align and integrate behavior and semantic representations. Extensive experiments validate the superiority of our method across various scenarios.

% In this paper, we propose \textbf{SIGER},  a framework designed to improve the modeling of user preferences by uncovering more comprehensive semantic relationships between items. Specifically, we first construct an Enhanced Item-Item Semantic Graph by integrating collaborative signals extracted from user-item interactions with item-item similarity semantic signals derived from raw modality features. This graph enables the learning of richer item semantic representations. To mitigate the impact on semantic representations of structural noise in the semantic graph caused by low-quality raw modality features, we propose a modulus-based personalized embedding perturbation mechanism. This mechanism generates different perturbed semantic representation views and leverages contrastive learning to derive stable and robust semantic representations. Finally, we develop a dual representation alignment mechanism to align and further fuse behavior and semantic representations. Extensive experiments on four datasets demonstrate the effectiveness of the proposed approach in various recommendation scenarios.

\newpage
\bibliography{aaai2026}

\end{document}